\documentclass{article}

\usepackage{cite}
\usepackage{arxiv}

\usepackage[utf8]{inputenc} % allow utf-8 input
\usepackage[T1]{fontenc}    % use 8-bit T1 fonts
\usepackage{hyperref}       % hyperlinks
\usepackage{url}            % simple URL typesetting
\usepackage{booktabs}       % professional-quality tables
\usepackage{amsfonts}       % blackboard math symbols
\usepackage{nicefrac}       % compact symbols for 1/2, etc.
\usepackage{microtype}      % microtypography
\usepackage{lipsum}
\usepackage{graphicx}
\usepackage{subcaption}
\usepackage{amsmath}
\usepackage{amssymb}
\usepackage{commath}
\usepackage{authblk}

\title{Intelligent Surveillance of World Health Organization (WHO) Integrated Disease Surveillance and Response (IDSR) Data in Cameroon Using Multivariate Cross-Correlation}

\author[1, 2]{ Jianzhi Liu \thanks{Equal contribution}}
\author[1, 2]{ Ziming Yang \thanks{Equal contribution}}
\author[1]{ Jesse E. Engelberg}
\author[1, 4,6]{Frankline S. Nsai}
\author[5]{Serge Bataliack}
\author[1,2,3]{Vikash Singh \thanks{Correspondence to vrs26@cam.ac.uk}}

\affil[1]{\footnotesize Datareach LLC, Los Angeles, USA}
\affil[2]{\footnotesize University of California, Los Angeles, Los Angeles, USA}
\affil[3]{\footnotesize University of Cambridge, Cambridge, UK}
\affil[4]{\footnotesize Songhai Labs, Yaounde, Cameroon}
\affil[5]{\footnotesize World Health Organization, Yaounde, Cameroon}
\affil[6]{\footnotesize University of Buea, Buea, Cameroon}

\begin{document}
\maketitle

\begin{abstract}
\textbf{Background:} As developing countries continue to face challenges associated with infectious diseases, the need to improve infrastructure to systematically collect data which can be used to understand their outbreak patterns becomes more critical. The World Health Organization (WHO) Integrated Disease Surveillance and Response (IDSR) strategy seeks to drive the systematic collection of surveillance data to strengthen district-level reporting and to translate them into public health actions \cite{world2000integrated}. Since the analysis of this surveillance data at the central levels of government in many developing nations has traditionally not included advanced analytics, there are opportunities for the development and exploration of computational approaches that can provide proactive insights and improve general health outcomes of infectious disease outbreaks. 

\textbf{Methods:} We propose a multivariate time series cross-correlation analysis as a foundational step towards gaining insight on infectious disease patterns via the pairwise computation of weighted cross-correlation scores for a specified disease across different health districts. Following the computation of weighted cross-correlation scores, we apply an anomaly detection algorithm to assess how outbreak alarm patterns align in highly correlated health districts (regions). Our methodology is compatible with the surveillance data collected via the IDSR strategy and the data used for this analysis was provided in collaboration with officials from the WHO Country Office in Cameroon, the Cameroon Ministry of Public Health, and Songhai Labs. More specifically, the analysis was performed using surveillance data for infectious diseases in Cameroon between the years of 2014 and 2017. 

\textbf{Result:} We demonstrate how multivariate cross-correlation analysis of weekly surveillance data can provide insight into infectious disease incidence patterns in Cameroon by identifying highly correlated health districts for a given disease, using malaria as an initial proof of concept. We further demonstrate scenarios in which identification of highly correlated districts aligns with alarms flagged using a standard anomaly detection algorithm, hinting at the potential of end to end solutions combining anomaly detection algorithms for flagging alarms in combination with multivariate cross-correlation analysis. 

\textbf{Conclusion:} We present a methodology for providing insights on infectious disease surveillance data based on multivariate cross-correlation analysis that is compatible with the data collection efforts currently being facilitated by the WHO IDSR surveillance program. We demonstrate the methodology on real-world surveillance data for infectious diseases in Cameroon from the years 2014-2017 and deduce correlations between health districts with respect to a particular infectious disease. We use the IDSR data of Cameroon as a case study, and hope that future work can expand on these approaches to better analyze surveillance data in a way that is useful in outbreak mitigation efforts. We are the first group to apply advanced analytics towards this data, and the methodology demonstrated in this paper is more generally a step towards bridging the gap between research on computational methods for analysis of surveillance data and real-world  surveillance of infectious diseases, especially in resource-limited scenarios associated with surveillance efforts in many developing countries.  
\end{abstract}

% keywords can be removed

\section{Background}
\subsection{Surveillance of Infectious Diseases}
Infectious diseases have long plagued developing countries, which often do not have the infrastructure and resources needed for effective eradication efforts \cite{morens2004challenge, lederberg2003microbial, binder1999emerging, amarasinghe2011dengue}. The need for a standardized infrastructure for data collection is highlighted by the fact that there is a large variance within individual nations in terms of ability to conduct surveillance efforts \cite{nsubuga2010strengthening, franco2006improving, sow2010trained}. In 1998 the World Health Organization Regional Office for Africa (WHO/AFRO), began to promote the Integrated Disease Surveillance and Response (IDSR) framework across all its members \cite{nsubuga2010implementing, perry2007planning, kabore2001technical, world1999regional}. The WHO IDSR framework paves the way towards realizing core capacities of International Health Regulations (IHR) to detect, assess, and respond to all events that may constitute public health emergencies of international concern (PHEICs) and report them to the WHO \cite{kasolo2013idsr}. Laboratory capacities, communication systems, logistics, and insufficient trained personnel are critical obstacles to successful implementation of IDSR \cite{phalkey2013challenges} \cite{sow2010trained} \cite{kebede2010trends}. Despite challenges in IDSR implementation, this led to an increase in surveillance data collection and created the opportunity to translate this surveillance data into actionable insights. Cameroon adopted the IDSR framework in 2003, recording district-level infectious disease counts on a weekly basis and forwarding them to the Ministry of Public Health for analysis. Figure 1 (taken from \cite{ngwa2015cholera}) provides a graphical overview of how surveillance data is collected in line with the WHO IDSR framework, beginning with data collection at the community/health district levels and final analysis at the central level. The IDSR framework in Cameroon has raised the consistency of district level data collection, but translation of this surveillance data into actionable insights has not been as successful. Additionally, although statistical techniques for time series analysis have existed for many years, they have not yet been widely applied to surveillance data collected through the WHO IDSR framework to drive actionable insights. 

\begin{figure}[h]
\includegraphics[width=1\textwidth]{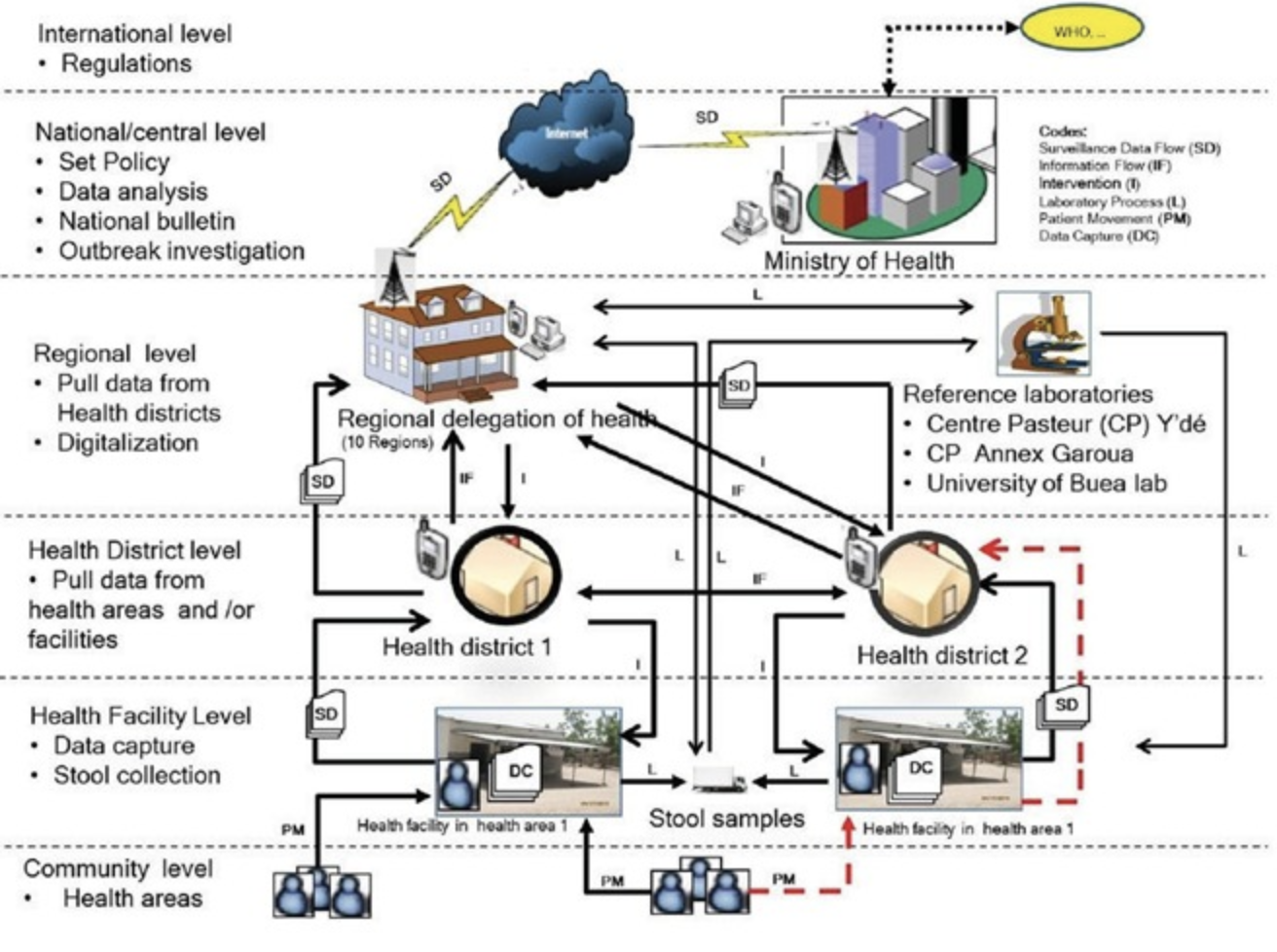}
\caption{Graphic of Surveillance Disease Data Collection Infrastructure in Cameroon from \cite{ngwa2015cholera}}
\end{figure}

\subsection{Cross-Correlation Between Time Series}
In times series analysis, the cross-correlation of a pair of time-series is the correlation between the pair at different times, as a function of the different times. Let $(X_t, Y_t)$ be a pair of random processes and t be any point in time. The cross-correlation between X and Y at times $t_1$ and $t_2$ respectively is defined by the equation below:

\begin{equation}
R_{XY}(t_1, t_2) = \mathbb{E}[X_{t1}\overline{Y_{t2}}]
\end{equation}

\subsection{Related Work}

The statistical analysis of disease surveillance data has been a topic of academic study for many years. The methods proposed often use standard surveillance datasets which are often higher in sample size and quality than those available from surveillance data collected in low resource settings, and utilize additional information. Many efforts have focused on model-based efforts for multivariate time series surveillance data rather than analysis geared towards explicitly identifying shared patterns in infectious disease surveillance counts between health districts. 

A nonlinear model for the analysis of multivariate time series data, with region-specific random effects to address varying transmission of a pathogens across regions has been shown \cite{paul2011predictive}. The model utilizes monthly counts of meningococcal disease cases in 94 departments of France (excluding Corsica) and weekly counts of influenza cases in 140 administrative districts of Southern Germany, and compare performance via one-step-ahead predictions. 

 A framework for the statistical analysis from counts of infectious diseases which is based on Poisson branching process model with immigration has been proposed \cite{held2005statistical}. The authors additionally propose a multivariate formulation, which has the ability to capture patterns in the spatial spread of a disease over time. Another approach involves a multivariate time series model based approach that is enhanced to deal with possible dependence between different disease counts from different pathogens \cite{paul2008multivariate}. The results are demonstrated on weekly influenza and meningococcal disease counts from Germany and on influenza in the U.S.A. from 1996–2006, with air traffic information being integrated with the influenza data to provide information on the global dispersal of the pathogen into the model. The open source R package surveillance provides functionality for implementing statistical regression frameworks that can be applied to surveillance data \cite{meyer2014spatio}.  

One proposed methodology is using ARIMA (autoregressive, integrated, moving average) forecasting of infectious disease counts using surveillance data \cite{allard1998use}. Another study investigated the relationship between joining the World Trade Organization (WTO) and the availability of several commodities with both harmful and protective effects related to the development of noncommunicable disease using a comparative interrupted time-series analysis \cite{cowling2019world}. 

\section{Methods}
    \subsection{Central Idea}
    The core idea underlying the application of the multivariate cross-correlation based methodology presented here is straightforward: the correlations between time series of disease incidence in geographic regions can potentially provide insight on disease spreading patterns. Although such an analysis does not allow for inference of causality due to the number of confounding variables, it provides signals which can warrant deeper investigation from the public health perspective. More specifically, deeper analysis can be performed to understand how prior knowledge of variables involved in the spread of infectious disease patterns can explain the observed patterns and thus inform concrete mitigation strategies.  
    
    \subsection{Methodology}
    Our computational approach for intelligent surveillance consists of three parts: a) obtaining and processing surveillance data of weekly disease counts, b) computing the multivariate cross-correlation between pairs of regions, and c) identifying highly correlated regions for a specific disease to gain insights on its spreading pattern. 
    
    To obtain the WHO IDSR surveillance data on infectious disease incidence in Cameroon, we established a collaboration with officials from the WHO Country Office in Cameroon, the Ministry of Health of Cameroon, and Songhai Labs, a Cameroonian Knowledge Broker Platform.  The partnership was built through video calls, letters, and in-person meetings with health officials.  The data represents weekly counts of infectious diseases, in 189 different health districts in Cameroon for the years 2014 to 2017.
   
    From this data we extract the weekly counts of a disease D in $N$ different regions. Wherever a data point is missing in the time series, we use interpolation to fill in missing values with averages of the first available preceding and following entries. In cases in which the first or last entry of the time series is missing, we use the nearest point as an approximation. After data wrangling, we get $N$ number of time series, $\mathbf{t}_1, \mathbf{t}_2, \dots, \mathbf{t}_N$. For $i \in \{1,\dots,N\}$, $\mathbf{t}_i$ is the weekly counts of disease D in region $R_i$.
    
     We compute a correlation score for every pair of regions in Cameroon using the cross-correlation function ($ccf$) and a score function $s$. More complex methods for determining correlation between time series exist, however the $ccf$ serves as a simple standard starting point for such analyses. For quantifying general similarity in patterns between time series (including seasonality) the standard ccf is suitable, although the methodology can be repeated following pre-whitening of each time series to remove seasonality and auto-correlation in order to avoid 'spurious correlation' \cite{dean2016dangers, chatfield2003analysis}. For all $i, j \in \{1,\dots,N\}$ and $i \neq j$, $(R_i, R_j)$ is a region pair, and $\mathbf{t}_i$ and $\mathbf{t}_j$ are the time series for region $R_i$ and $R_j$ respectively. We adopt the $ccf$ function interface provided by the tseries \cite{ccfdocumentation} library in R. This implementation of $ccf$ takes as input two time series, $\mathbf{t}_1$ and $\mathbf{t}_2$, as well as a $lag$ parameter, and outputs a vector containing correlations of $\mathbf{t}_1$ and $\mathbf{t}_2$ from under $lag$-week lag to under $(-lag)$-week. Applying $ccf$ function to $\mathbf{t}_i$ and $\mathbf{t}_j$ with the parameter $lag=5$, we obtained the correlation vector $\mathbf{c}_{ij} \in \mathbb{R}^{11}$:
    \begin{equation} \label{corrVecDef}
        \mathbf{c}_{ij} = ccf(\mathbf{t}_i,\mathbf{t}_j,lag=5)
    \end{equation}
    For all $k \in \{-5,-4,\dots,0,\dots,4,5\}$, denote with ${c_{ij}}_k$ the correlation between $\mathbf{t}_i$ and $\mathbf{t}_j$ under $k$-week lag, where a positive $k$ means moving $\mathbf{t}_i$ ahead of $\mathbf{t}_j$ by $\abs{k}$ weeks, and a negative $k$ moving $\mathbf{t}_j$ ahead of $\mathbf{t}_i$ by $\abs{k}$ weeks. According to this definition, \begin{equation} \label{symmProp}
        {c_{ij}}_k = {c_{ji}}_{-k}
    \end{equation}
    Following Equation \ref{corrVecDef} and the definition of ${c_{ij}}_k$,
    \begin{equation}
        \mathbf{c}_{ij} = [{c_{ij}}_{-5},\dots,{c_{ij}}_0,\dots,{c_{ij}}_5]^T
    \end{equation}
    We define the score function $s:\mathbb{R}^{11}\rightarrow\mathbb{R}$ as
    \begin{equation}
        s(\mathbf{c}) = \sum_{k=-5}^{5} \frac{10}{\abs{k}+1} c_k
    \end{equation}
    The score function is the weighted sum of all the entries in the correlation vector $\mathbf{c}_{ij}$. The weight assigned to the correlation under $k$-week lag is stipulated as $\frac{10}{\abs{k}+1}$. The weight given to the correlation vector is inversely correlated with the absolute value of the lag, intuitively placing higher importance on closer times. 
    Eventually, the correlation score $cs$ for $(R_i, R_j)$ is
    \begin{equation} \label{csDef}
        cs_{ij} = s(\mathbf{c}_{ij})
    \end{equation}
    Applying the computation outlined above for all regional pairs for a given disease, we obtain a set of corresponding correlation scores.
    \begin{equation}
        \mathbb{S} = \{ cs_{ij} \in \mathbb{R} \mid i,j = 1,2,\dots,N, i \neq j \}
    \end{equation}
    Importantly, the metric $cs$ is relative, with insight derived only in comparison with other regions, and is designed specifically with the goal of identification of positive trends between two time series. A negative correlation score between two regions indicates that when one region experiences an outbreak, the disease incidence in the other region tends to drop, which is not the trend of interest when using these scores to provide insight on the ramifications of an initial outbreak. 
    
  To evaluate our methods we examined regions with high correlation scores to see how their alarm patterns aligned using an anomaly detection algorithm. We used the EARS (Early Aberration Detection System) C method for selected region pairs \cite{salmon2016monitoring}. This anomaly detection method computes a threshold (prediction interval) for the number of counts at each time step based on values from the recent past. If the number of counts for a step is higher than a threshold, an alarm is raised. Specifically for the threshold, we calculate an approximate (two-sided) $(1 - \alpha)\%$ prediction interval, based on the assumption that the difference between the expected value and the observed value divided by the standard derivation of counts over the sliding window, called $C$, follows a standard normal distribution in the absence of outbreaks.
  
  We utilized the computed correlation scores, $\mathbb{S}$, in tandem with the results of the EARS algorithm to identify how high weighted correlation scores aligned with alarms flagged using the EARS algorithm. Once we mark a region $R_i$ as experiencing an outbreak, we obtain the subset $\mathbb{S}_i$ of $\mathbb{S}$, the set of all the correlation scores of region pairs involving $R_i$.
    \begin{equation}
        \mathbb{S}_i = \{ cs_{ij} \in \mathbb{R} \mid j = 1,\dots,i-1,i+1,\dots,N \}
    \end{equation}
  In practice the regions with the highest correlation scores can be further investigated and flagged for warning as a result of the initial outbreak. While extrapolating causal relationships is difficult for such time series due to the multitude of factors that contribute, historical analysis of consistent infectious disease correlation patterns can create foundations for intelligent warning systems since these correlations inherently capture information about the nature of how infectious disease outbreaks occur. 
  
 \section{Results}
    \subsection{Inter-regional Correlation Analysis}
    We show surveillance data of malaria as an example to demonstrate our methodology. Table 1 shows the format of the data: the $i$th row represents the time series $\mathbf{t}_i$ of region $R_i$. Weeks are marked consecutively W1, W2, ..., W207. The numbers under column W$i$ are the reported counts of Malaria cases in different regions during the corresponding week.
    
    Our analysis begins with a set of time series $\{\mathbf{t}_i \mid i = 1,2,\dots,N\}$ as represented in Table 1.

    \begin{figure}[h]
        \centering
        \includegraphics[width=1\textwidth]{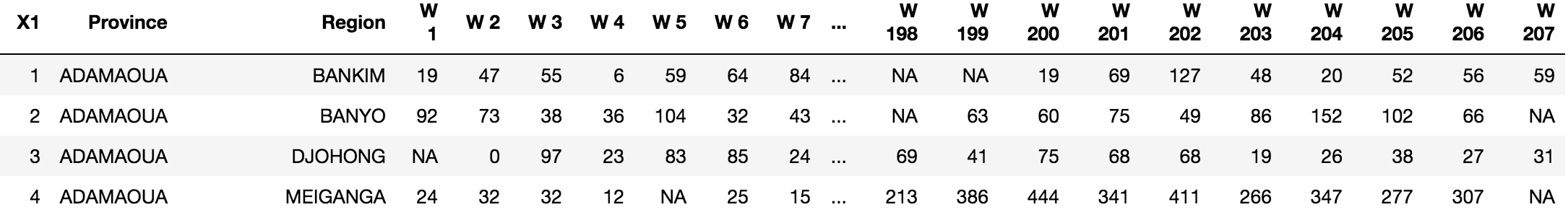}
        \caption{A Snippet of Malaria Surveillance Data}
    \end{figure}

    \begin{figure}[h]
        \centering
        \begin{subfigure}[h]{0.45\textwidth}
            \includegraphics[width=1\textwidth]{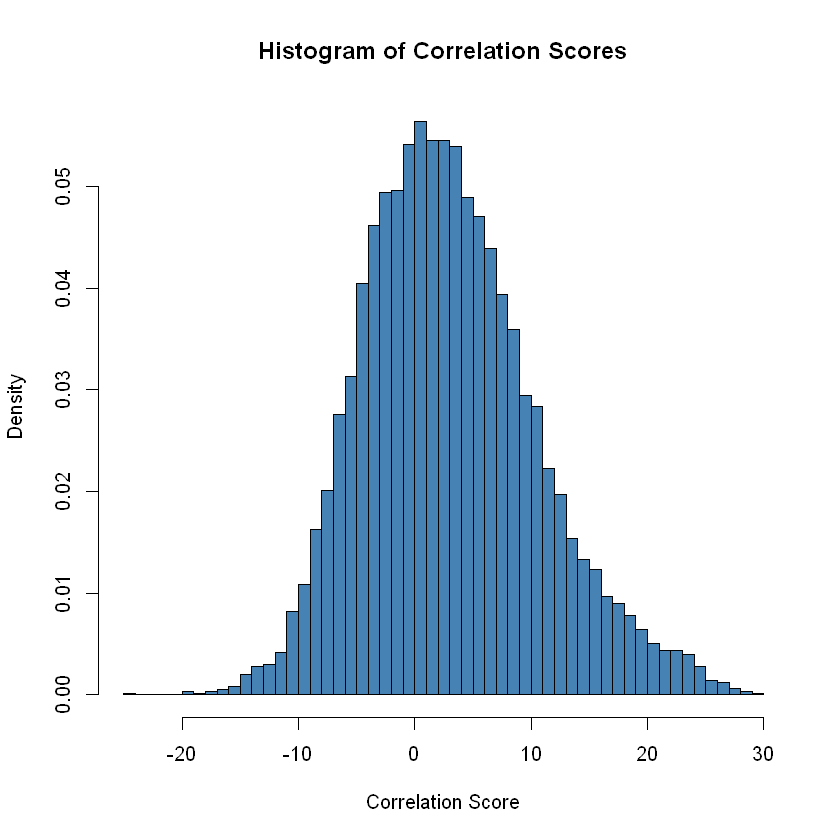}
            \caption{Distribution: Mean=3.02, SD=7.44; Min=-24.22, Median=2.40, Max=29.96}
        \end{subfigure}
        \begin{subfigure}[h]{0.45\textwidth}
            \includegraphics[width=1\textwidth]{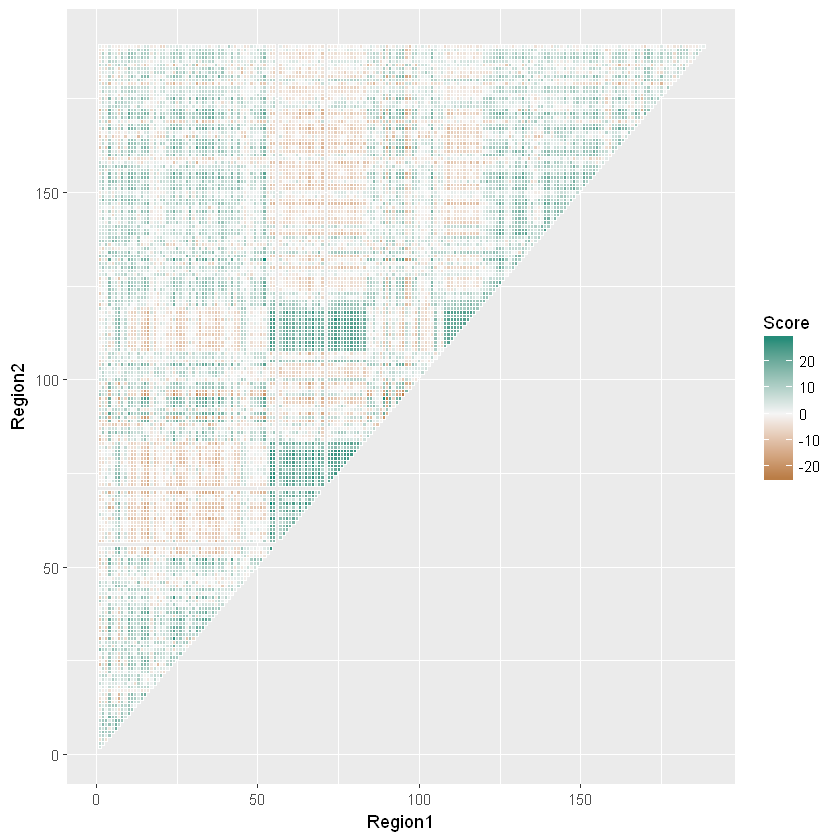}
            \caption{Heatmap of correlation scores. Strong green colors indicate regions that are associated, however proximity of indices does not reflect proximity in geographical distance}
        \end{subfigure}
        \caption{Distribution and Heatmap of Correlation Scores in $\mathbb{S}^{'}$}
    \end{figure}
    
    Figure 3a shows the histogram of correlation scores for malaria between different regions, which are normally distributed with a slight skew. We previously calculate $\mathbb{S}$, the set of correlation scores for all possible region pairs. According to Equation \ref{symmProp} and \ref{csDef}, $\forall i, j \in \{1,\dots,N\}, i \neq j,$
    \begin{equation}
        cs_{ij} = 
        \sum_{k=-5}^{5} \frac{10}{\abs{k}+1} {c_{ij}}_k = 
        \sum_{k=-5}^{5} \frac{10}{\abs{-k}+1} {c_{ji}}_{-k} =
        \sum_{{-k}=-5}^{5} \frac{10}{\abs{-k}+1} {c_{ji}}_{-k} =
        cs_{ji}
    \end{equation}
    Thus, we only compute a subset of $\mathbb{S}$:
    \begin{equation}
        \mathbb{S}^{'} = \{ cs_{ij} \in \mathbb{R} \mid i,j = 1,2,\dots,N, i < j \}
    \end{equation}
    Figure 3b presents a heatmap of the correlation scores in $\mathbb{S}^{'}$.  The heatmap shows which region pairs are strongly positively correlated (green), weakly correlated (white) and strongly negatively correlated (brown). The correlations represent the average correlation for the entire time period.  Each grid in the heatmap represents a region pair, indicated by the indices on horizontal and vertical axes. Indices that are next to each other are close geographically.
 
  When using historical correlation patterns to provide insight on the potential future spreading patterns of infectious disease, it is critical to have an understanding on the stability of such patterns over time. In Figure 4, we compare heatmaps between two years period (2014-15 and 2016-2017) with each other the heatmap for the entire four year period (2014-2017). The correlation heatmap does not noticeably change between 2014-2015 and 2016-2017, and the two heatmaps produce almost the same result as the one comparing 2014 and 2017 (Figure 3). Just as in the 2014-2017 heatmap, where regions with indices from 60 to 80 are strongly correlated, the same pattern exists in both 2014-2015 and 2016-2017, meaning that these regions have preserved their correlation pattern through these years.  We note that some regions (such as in the upper right) do change, however the major patterns remain intact.
    
    \begin{figure}[h!]
        \begin{subfigure}[h]{0.35\textwidth}
            \includegraphics[width=1\textwidth]{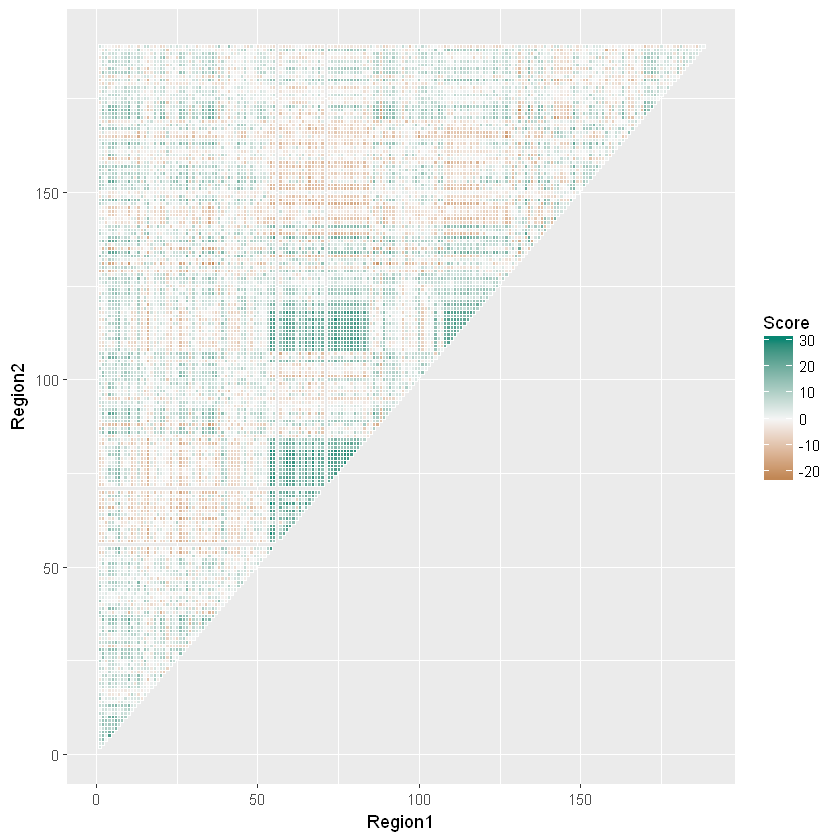}
            \caption{14-15}
        \end{subfigure}
        \begin{subfigure}[h]{0.35\textwidth}
            \includegraphics[width=1\textwidth]{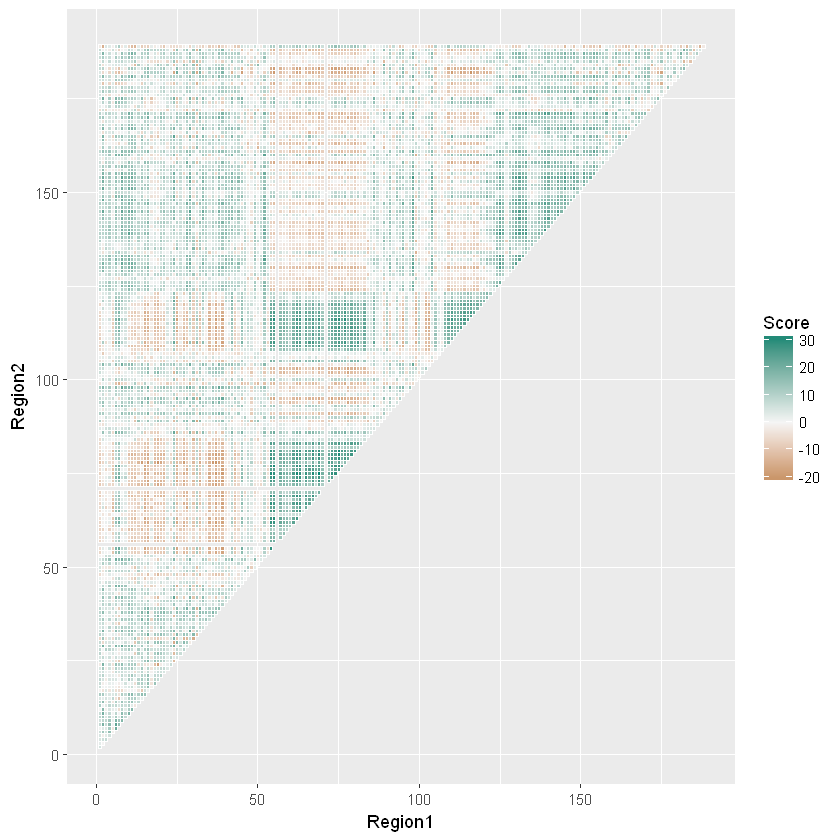}
            \caption{16-17}
        \end{subfigure}
        \begin{subfigure}[h]{0.35\textwidth}
            \includegraphics[width=1\textwidth]{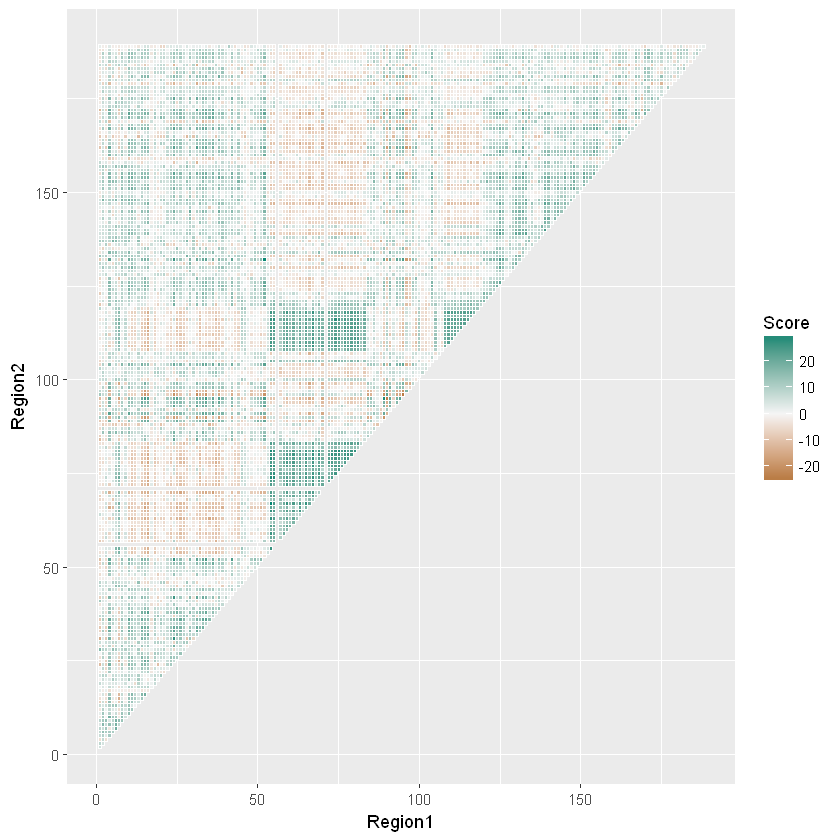}
            \caption{14-17}
        \end{subfigure}   
         \caption{Heatmap of Correlation Scores between Every Two Regions in Years 2014-2015, 2016-2017, 2014-2017}
    \end{figure}

    \begin{figure}[h!]
        \centering
        \begin{subfigure}[h]{0.3\textwidth}
            \includegraphics[width=1\textwidth]{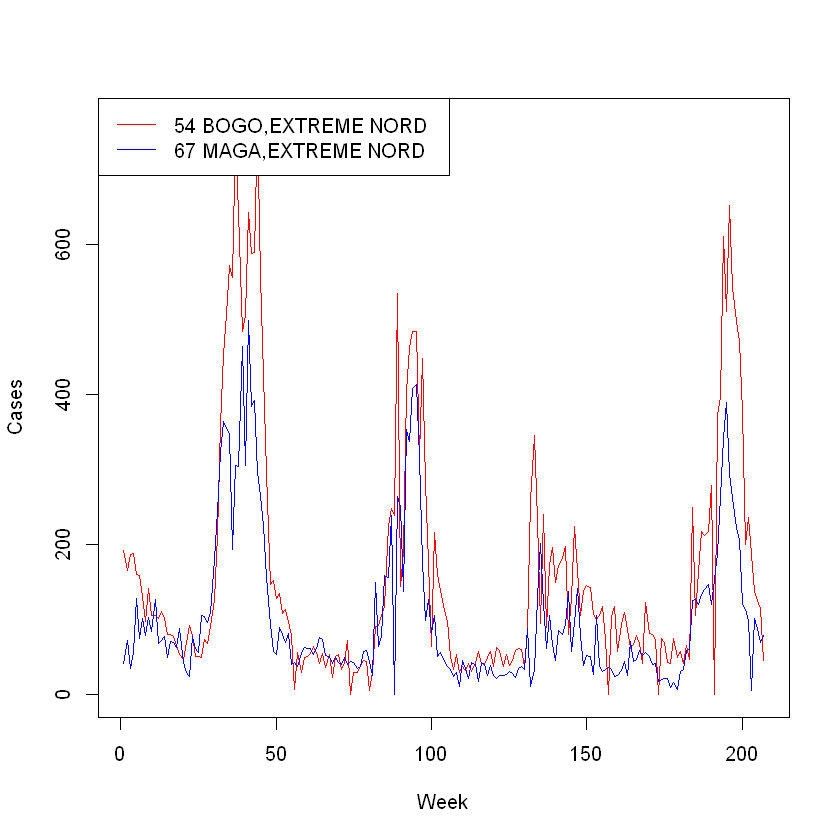}
        \end{subfigure}
        \begin{subfigure}[h]{0.3\textwidth}
            \includegraphics[width=1\textwidth]{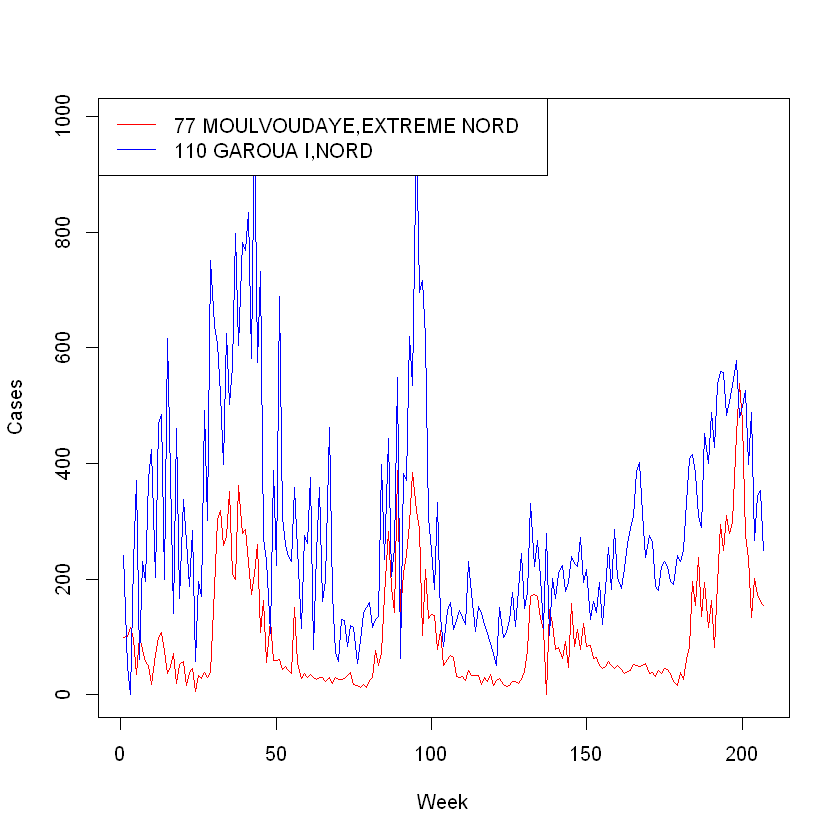}
        \end{subfigure}
        \begin{subfigure}[h]{0.3\textwidth}
            \includegraphics[width=1\textwidth]{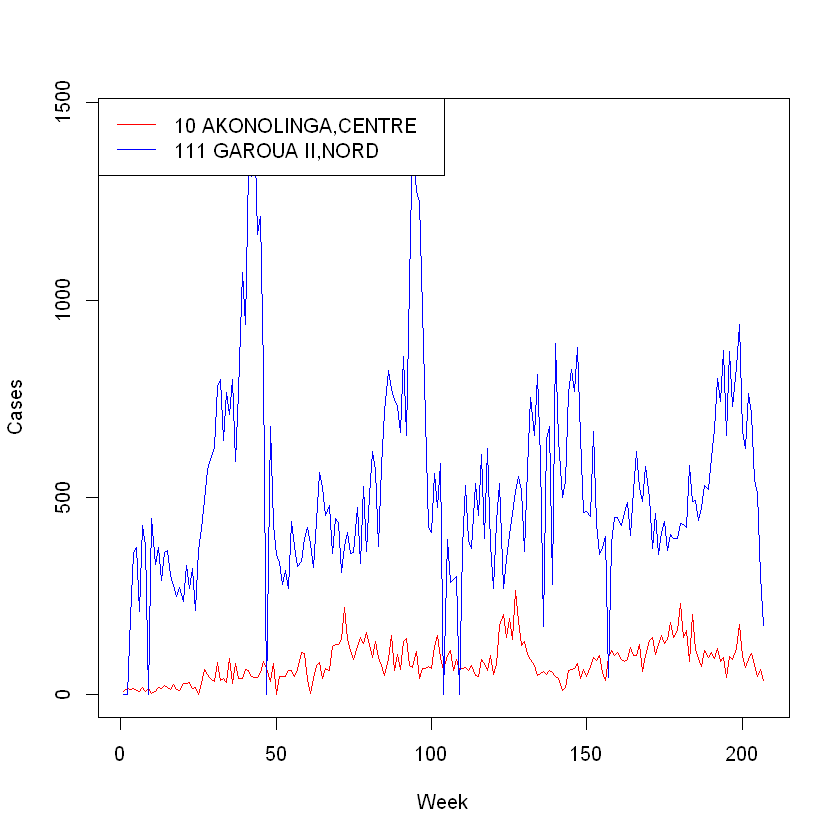}
        \end{subfigure}
        \begin{subfigure}[h]{0.3\textwidth}
            \includegraphics[width=1\textwidth]{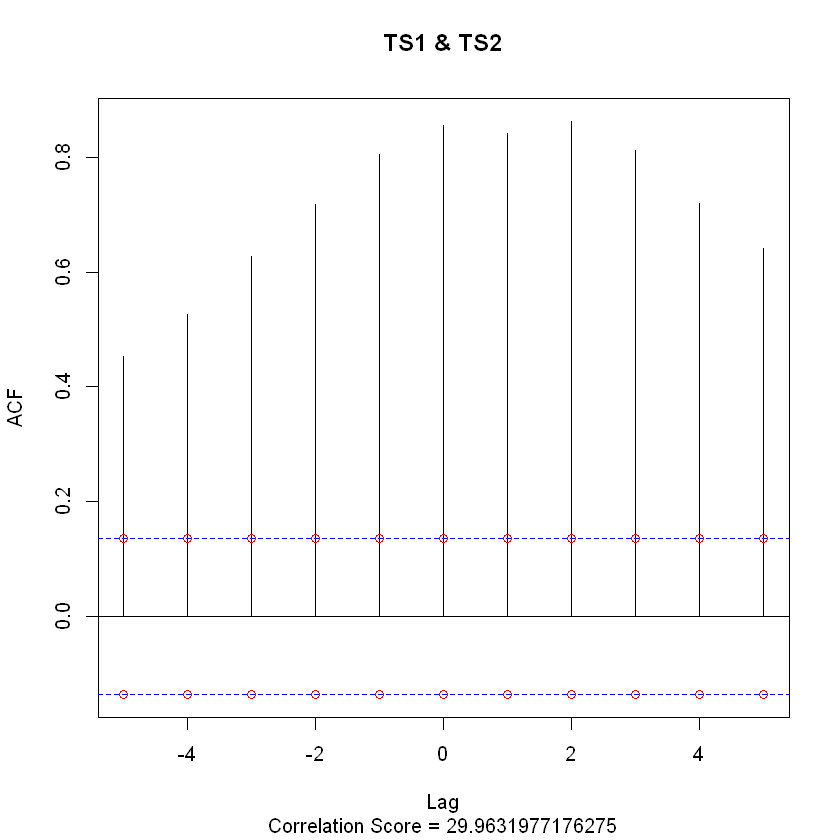}
            \caption{$(R_{54},R_{67})$}
        \end{subfigure}
        \begin{subfigure}[h]{0.3\textwidth}
            \includegraphics[width=1\textwidth]{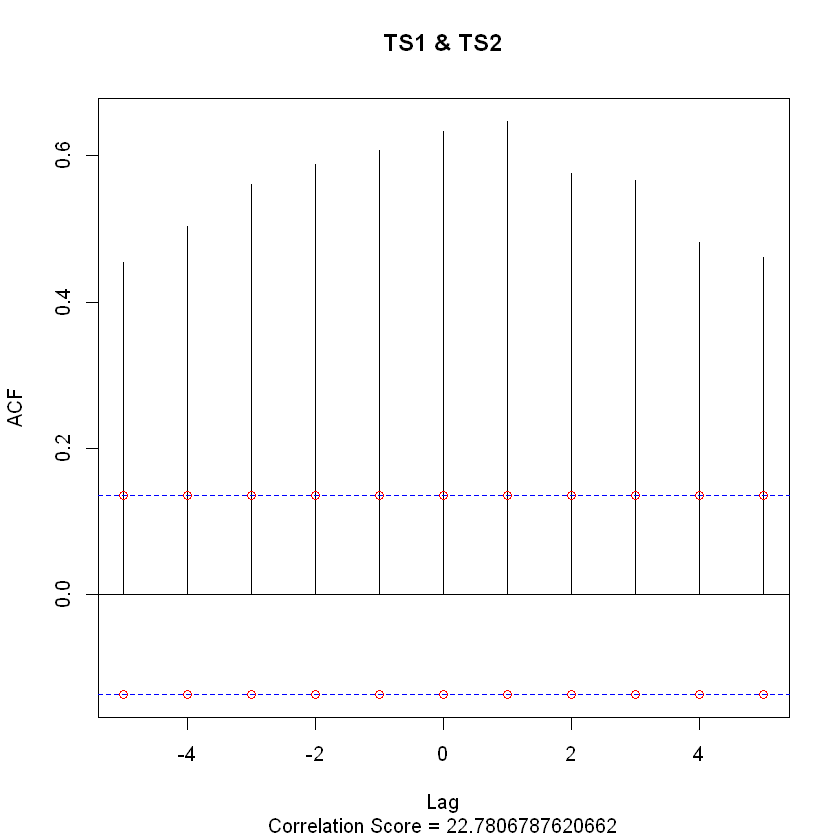}
            \caption{$(R_{77},R_{110})$}
        \end{subfigure}
        \begin{subfigure}[h]{0.3\textwidth}
            \includegraphics[width=1\textwidth]{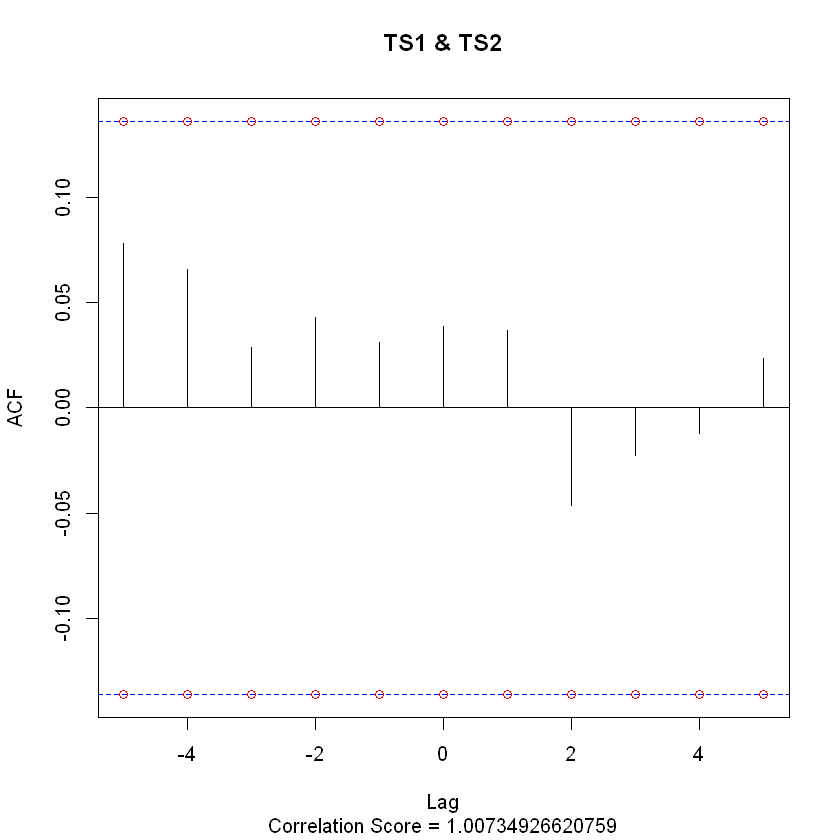}
            \caption{$(R_{10},R_{111})$}
        \end{subfigure}
        \caption{Examples of Time Series and Cross-Correlation Plots}
    \end{figure}
    
\subsection{Evaluation}
    
    In order to further validate our method, we analyzed the correlation scores of several specific region pairs. The highest observed correlation score $29.96$, comes from the region pair $(R_{54},R_{67})$ representing the Bogo and Maga districts, which are next to each other in the Far North region. The time series $\mathbf{t}_{54}$ and $\mathbf{t}_{67}$ are plotted in Figure 5a along with their correlation vector $\mathbf{c}_{54,67}$. From the plot of the time series, it is clear that $\mathbf{t}_{54}$ and $\mathbf{t}_{64}$ are highly correlated. Moreover, in the correlation vector score plot where the blue dotted line is the threshold for the 0.05 significance level, under lags of various magnitude, the correlations of $\mathbf{t}_{54}$ and $\mathbf{t}_{64}$ computed by the $ccf$ function are always well above that threshold. Both plots support the high correlation score of $(R_{54},R_{67})$ and suggest strong association.

    Another region pair, $(R_{77}, R_{110})$ (Moulvoldaye and Garoua districts, which are not adjacent but relatively close together) has a relatively high correlation score $c_{54,67} = 22.78$. The corresponding plots are shown in Figure 5b. In the time series plot, $\mathbf{t}_{77}$ and $\mathbf{t}_{110}$ are somewhat correlated: there are time periods when they are in sync (from Week 25 to 40), and others where they deviate (from Week 160 to 170). The correlation of $(R_{77}, R_{110})$ is significant, but not as strong as that of $(R_{54}, R_{67})$.
    
    The region pair $(R_{10}, R_{111})$. $c_{10,111} = 1$, has a very low correlation score as shown in Figure 5c. There is very little observable correlation between $\mathbf{t}_{10}$ and $\mathbf{t}_{111}$, and the values within the cross-correlation plot are not measured to be significant.  This makes sense since these regions (Akonlinga and Garoua II), are far apart geographically and may not have a strong influence on each other in terms of respective malaria incidence patterns. 
    
    The correlation plots can be used to dig deeper into how an outbreak in one region has historically been associated with changes in incidence levels in other regions. Looking at the correlations with respect to specific lags can further provide more granular information on the time scale in which relevant patterns are observed, giving insight on the temporal aspect of the observed correlation scores.

    A practical use case of the described methodology is to analyze the weighted cross-correlation scores for a given region once it is known that an outbreak has taken place. To illustrate, assume that an anomaly detection algorithm has detected an anomaly in $R_{54}$, which indicates a possible outbreak. We compute the $k$ largest correlation scores in $\mathbb{S}_{54} = \{c_{54,j} \mid j = 1,\dots,53,55,\dots,N \}$, and the regions corresponding to those scores. In this case, we stipulated $k = 5$. The results are shown in Table 2 and serve as a sample output for this method given an input region which has experienced an alarm for an outbreak. 

    \begin{table}[h]
    \begin{tabular}{ |l|c| } 
    \hline
    \multicolumn{1}{|c|}{Region Pairs} & Correlation Scores \\ 
    \hline
    54 BOGO,EXTREME NORD; 67 MAGA,EXTREME NORD & 29.96 \\ 
    54 BOGO,EXTREME NORD; 81 TOKOMBERE,EXTREME NORD & 28.87 \\ 
    54 BOGO,EXTREME NORD; 79 PETTE,EXTREME NORD & 28.31 \\
    54 BOGO,EXTREME NORD; 74 MOGODE,EXTREME NORD & 28.25 \\
    54 BOGO,EXTREME NORD; 77 MOULVOUDAYE,EXTREME NORD & 27.56 \\
    \hline
    \end{tabular}
    \vspace{1em}
    \caption{Top 5 Correlation Scores in $\mathbb{S}_{54}$}
    \label{table:1}
    \end{table}
    \begin{figure}[h!]
        \centering
        \begin{subfigure}[h]{0.4\textwidth}
            \includegraphics[width=1\textwidth]{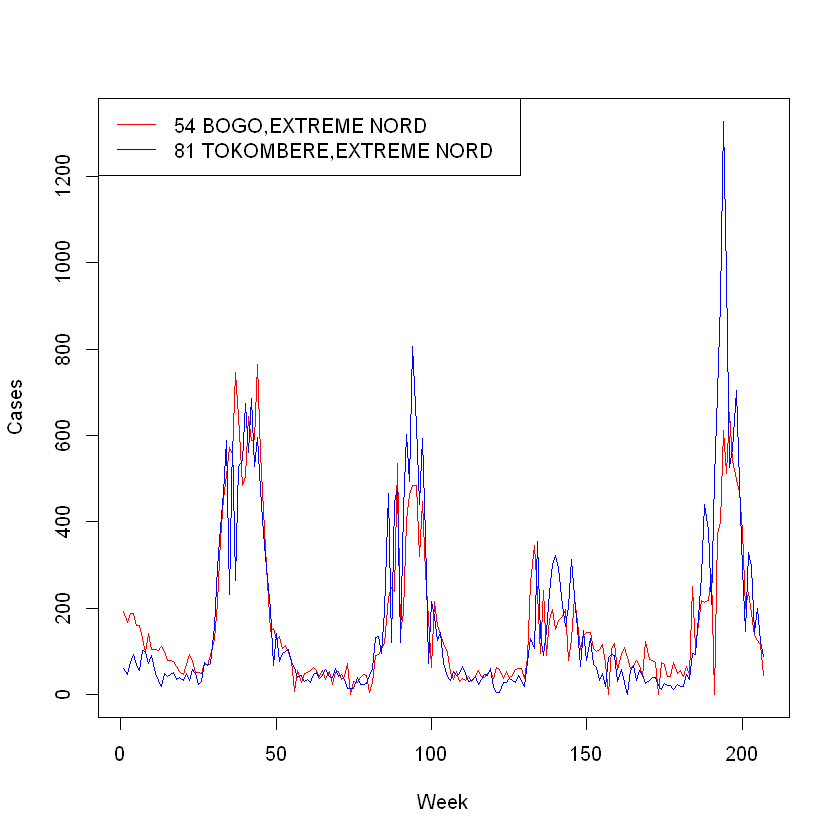}
        \end{subfigure}
        \begin{subfigure}[h]{0.4\textwidth}
            \includegraphics[width=1\textwidth]{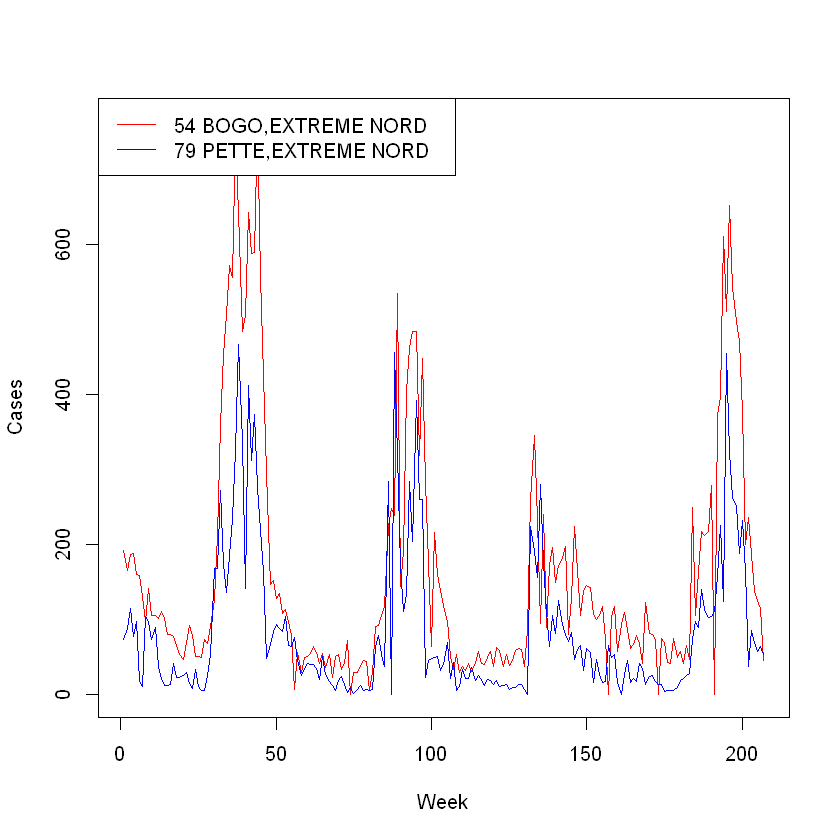}
        \end{subfigure}
        \begin{subfigure}[h]{0.4\textwidth}
            \includegraphics[width=1\textwidth]{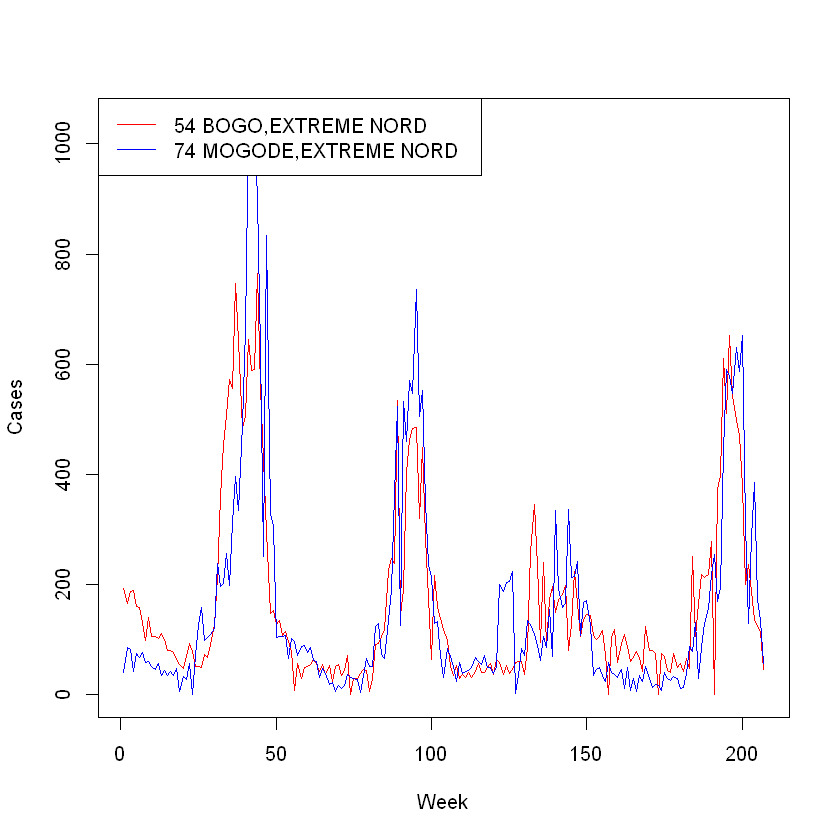}
        \end{subfigure}
        \begin{subfigure}[h]{0.4\textwidth}
            \includegraphics[width=1\textwidth]{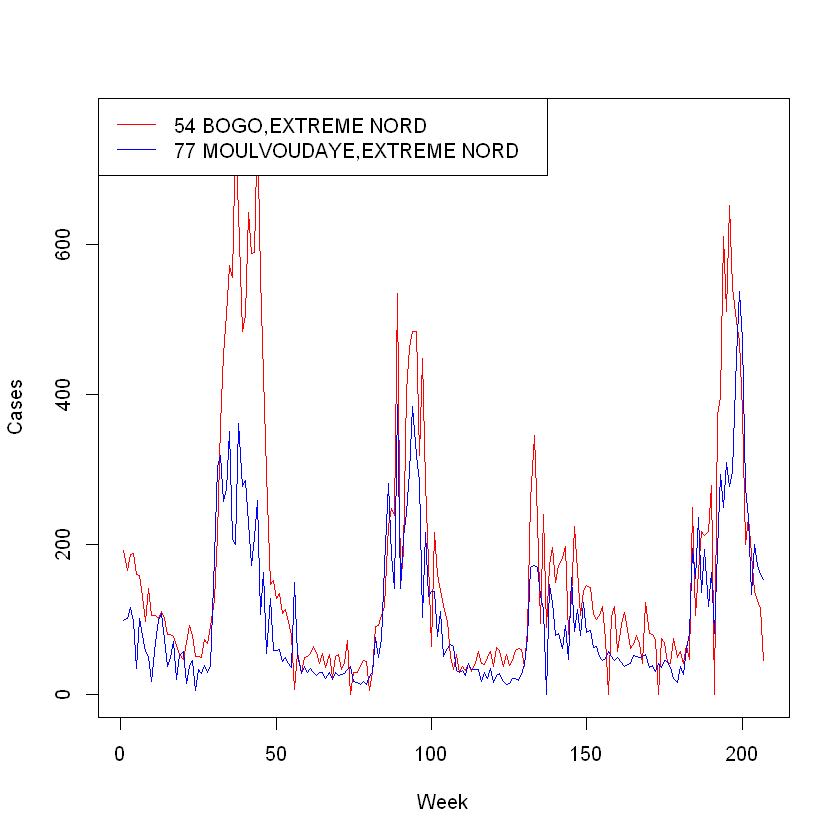}
        \end{subfigure}
    \iffalse
        \begin{subfigure}[h]{0.4\textwidth}
            \includegraphics[width=1\textwidth]{figures/step2c_2ccf.png}
            \caption{$(R_{54},R_{81})$}
        \end{subfigure}
        \begin{subfigure}[h]{0.4\textwidth}
            \includegraphics[width=1\textwidth]{figures/step2c_3ccf.png}
            \caption{$(R_{54},R_{79})$}
        \end{subfigure}
        \begin{subfigure}[h]{0.4\textwidth}
            \includegraphics[width=1\textwidth]{figures/step2c_4ccf.png}
            \caption{$(R_{54},R_{74})$}
        \end{subfigure}
        \begin{subfigure}[h]{0.4\textwidth}
            \includegraphics[width=1\textwidth]{figures/step2c_5ccf.png}
            \caption{$(R_{54},R_{77})$}
        \end{subfigure}
    \fi
        \caption{Time Series Corresponding to 4 Region Pairs Identified in Table 1}
    \end{figure}

    The two time series corresponding to $(R_{54},R_{67})$ are shown in Figure 5a, while the time series plots corresponding to the remaining entries of Table 1 can be found in Figure 6.  

    \begin{figure}[h!]
        \centering
        \begin{subfigure}[h]{0.45\textwidth}
            \includegraphics[width=1\textwidth]{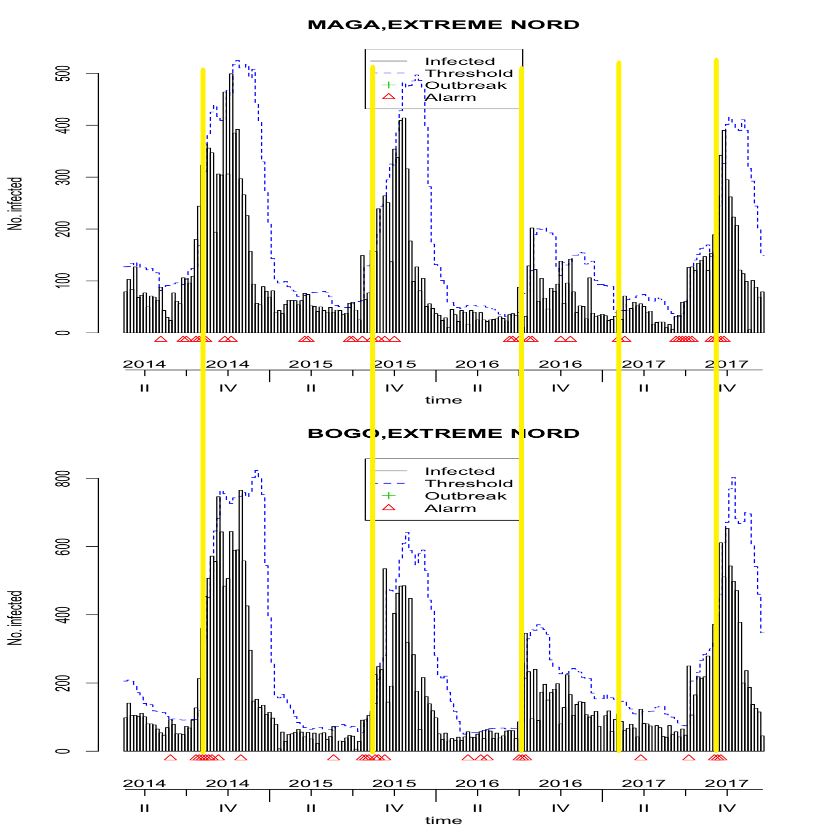}
        \end{subfigure}
        \begin{subfigure}[h]{0.45\textwidth}
            \includegraphics[width=1\textwidth]{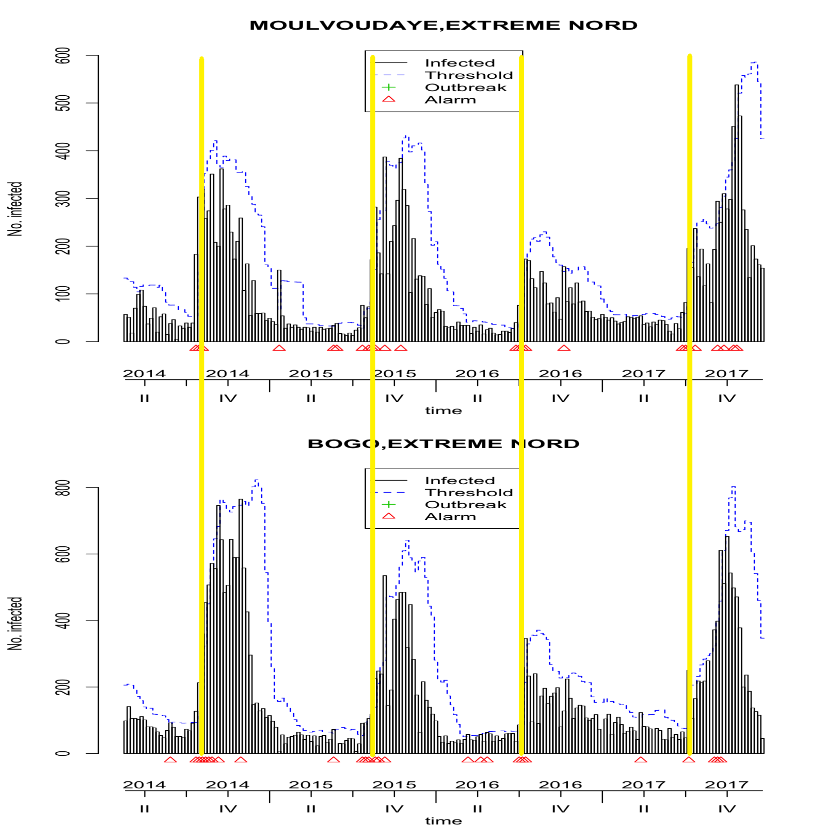}
        \end{subfigure}
        \begin{subfigure}[h]{0.45\textwidth}
            \includegraphics[width=1\textwidth]{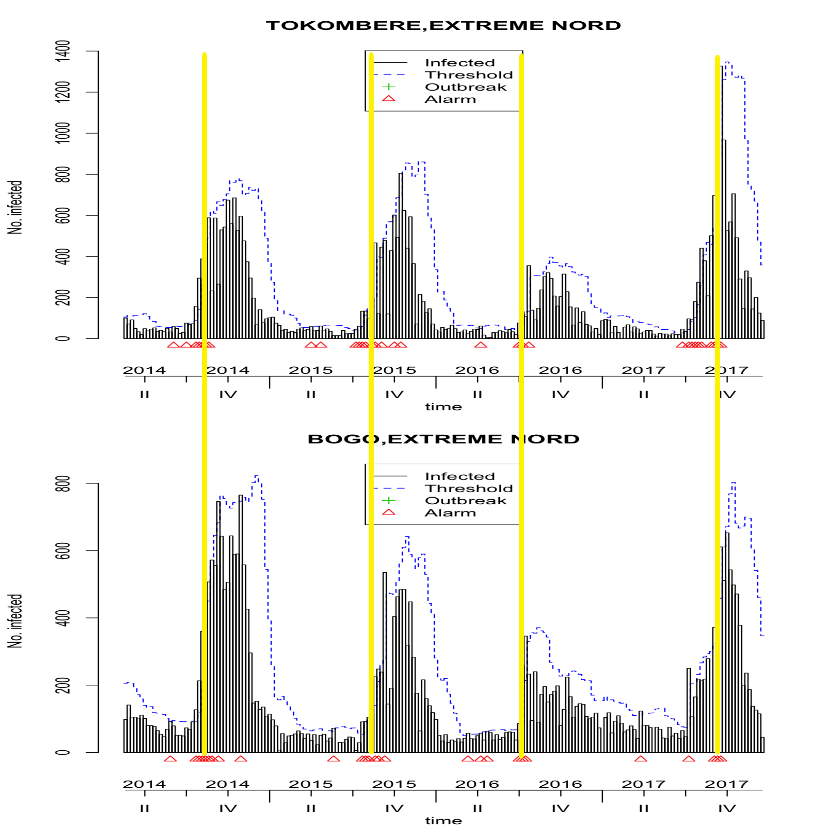}
        \end{subfigure}
         \caption{Outbreak Alarm Patterns Flagged using EARS C Anomaly Detection for 3 Region Pairs in Table 1}
    \end{figure} 

    We then apply the EARS C method on the pairs of regions that have the highest correlation scores to analyze how patterns in flagging alarms are aligned. For example, we take the pairs $R_{54}$ and $R_{77}$, $R_{54}$ and $R_{67}$, and $R_{54}$ and $R_{77}$ and compare their outbreak patterns using the EARS C method with a look-back period of 7 weeks. This method will detect whether the disease counts from a given week are abnormal based on disease counts in the previous 7 weeks. As evident from Figure 7, the two regions in each pair have very similar alarm patterns.
    
    Furthermore, these alarms appear clustered around the same time. The yellow lines on Figure 7 indicate alarm clusters, which overlap in correlated regions. Instances where alarms in one region precede another could indicate disease spreading from one region to another. We observed similar alarm clustering in other regions with high correlation scores, confirming the reliability of this method.

\section{Discussion}
    %\subsection{Practical Implications}
    The initial goal of the WHO IDSR strategy was to develop actionable insights to improve health outcomes of infectious disease outbreaks, but so far in Cameroon, analysis of surveillance data has been mainly descriptive. The lack of advanced analytics on surveillance data collected through the WHO IDSR strategy limits the ability of public health officials to improve their decision-making based on insights and trends from the collected data. The methodology introduced in this paper serves as a step towards bridging the gap between advanced analytics and surveillance data collected through the WHO IDSR strategy. 
    
    We demonstrate the potential for a weighted cross-correlation analysis in quantifying similarity in surveillance time series for an infectious disease. Although causal relationships can't be ascertained from analysis of multivariate cross correlation, general trends can be assessed in evaluating the similarity of disease frequency patterns across many health districts. The evaluated similarity in disease frequency between health districts is a reflection of numerous factors, including general seasonality and outbreak spreading patterns. We further demonstrated that using a weighted cross correlation metric has the potential to cluster similar regions based on similar alarms flagged by an outbreak detection algorithm (EARS C1). This signals utility in providing insight on how infectious diseases may spread and flagging health districts that may be susceptible to the ramifications of an initial known outbreak. Using known weighted cross-correlation scores for a given disease can allow for more proactive monitoring of disease outbreaks even before the development of more sophisticated software systems, and serves as a gate to deploy more sophisticated techniques in artificial intelligence which can be better leveraged as size of the data increases. Proactive monitoring and utilization of historical surveillance to provide insight on correlation in disease frequency in different health districts can inform more efficient outbreak mitigation strategies as well. 
    
    The trends revealed via multivariate cross correlation analysis be subsequently investigated to understand which factors may play a role in contributing to such trends. An analysis of such factors thereby may lead to a more mechanistic understanding of spreading between diseases, which can be leveraged by public health officials. For example, two highly correlated regions that are not necessarily close together geographically may share extensive modes of transportation. Investigating these modes of transportation and how they pertain to the spread of a certain infectious diseases can potentially lead to better preventative mechanisms for outbreak mitigation.

    %\subsection{Limitations and Extensions}
    One of the limitations of the current method is its use of the EARS C1 algorithm, which does not adequately capture the seasonality of disease outbreak patterns. As we can see from the outbreak patterns of different regions in Figure 7, there are clear seasonal trends in disease outbreaks. Specifically, the number of disease counts in the summer seems to be much higher than the average level every year, indicating a summertime peak period. Our current EARS C1 algorithm only considers a certain time period (7 weeks in this case), and thus misses seasonal trend. As a result, the algorithm tends to give unnecessary alerts when we are already on a high alert level. Investigating the use of anomaly detection algorithms that can take seasonal trends into account and understanding how the alarms cluster for highly correlated regions would be a natural extension to this work. 
    
    Another limitation is that pairwise cross-correlation analysis does not directly capture the dynamics of multiple simultaneous outbreaks. Exploring methodologies that can analyze these trends in the context of multiple simultaneous outbreaks would be an ideal avenue for future work. 
    
    It is important to note that the study only uses data over a four year period. As additional data collection infrastructure is put in place more surveillance data will become available for the continued development of methodologies for analysis. 
    For the purpose of demonstrating our methodology, this paper uses Malaria as an example to demonstrate a proof of concept, largely due to improved data reporting quality and quantity for Malaria specifically. While Malaria is spread indirectly through parasites, a the same approach could be used for a disease that is passed between humans and compare how the identified patterns differed. Since different diseases naturally have various of modes of transmission and hence different inter-regional spread pattern, future research could use multivariate cross-correlation to delve deeper into the differences in observed trends between types of diseases. 
    
\section{Conclusions}
WHO's Integrated Disease Surveillance and Response (IDSR) strategy has collected valuable disease surveillance data in Cameroon. Nevertheless, public health officials in Cameroon still face the challenge harnessing this information to obtain actionable insights. To issue this problem, we propose a computational methodology utilizing multivariate cross-correlation analysis to identify patterns of disease incidence between regions, which could in turn be used to drive more efficient mitigation efforts and drive investigation into mechanisms behind disease spread. In addition to presenting a mathematical formulation of this methodology, we present its application and analysis using real WHO surveillance data provided by officials from the WHO Country Office in Cameroon. Retroactive analysis, using malaria as an example, reveals the potential of the outlined methodology to provide strong signals for regional correlation of infectious diseases. As data collected through the WHO IDSR program increases in quantity and quality , computational approaches such as the one presented in this paper and future developed methods will become even more useful, ultimately improving disease mitigation and health outcomes. 

 \section{Acknowledgments}
This work was partially supported through the generous funding from the UCLA Global Citizens Fellowship Award to VS. 
\bibliography{references}
\nocite{*}
\bibliographystyle{unsrt}

\end{document}